\documentstyle[11pt,newpasp,twoside]{article}
\def\edcomment#1{\iffalse\marginpar{\raggedright\sl#1\/}\else\relax\fi}
\reversemarginpar

\begin{document}
\title{Finding out the Velocity Anisotropy Parameter for some Globular Clusters: the case of stationary model}
\author{Abdikul Ashurov}
\affil{Astronomy Department, Physical Faculty, National University of Uzbekistan,
Vuzgorodok, 700174 Tashkent, Uzbekistan}
\author{Salakhutdin Nuritdinov}
\affil{Astronomy Department, Physical Faculty, National University of Uzbekistan,
Vuzgorodok, 700174 Tashkent, Uzbekistan}

\begin{abstract}
The problem of determination of the velocity anisotropy parameter from star
density profiles in globular clusters is considered. The calculations are
performed for 10 globular clusters. The simple way for study of the dependence of
anisotropy parameter on radial distance is used.
\end{abstract}

\section{Introduction}

The anisotropic models occup an important place among various realistic models
for study of the dynamical evolution of globular clusters. In frame of this models it have
been work out the many evolution theories of the kinematics and dynamics of spherical
self-gravitating systems (e.g., Bettwieser \& Spurzem 1986; Spurzem 1996; Louis 1990;
Solanes \& Salvador-Sole 1990; Takahashi 1996). At the same time have been accumlated
many observational data. These data mostly concerned star counts and surface brightness
profiles, proper motions, stellar radial velocity dispersion, etc. (Meylan \& Heggie
1997). Despite of the presence such great quantity information, still there are
difficulties in comparision problem the observational data with results of numerical simulations
in the frame work of anisotropic models. One of causes of this phenomenon perhaps is concluded in
that for their performe we must have at least the value of the anisotropy parameter
$A=2-2\sigma_t^2 /\sigma_r^2$, where $\sigma_r^2$ and $\sigma_t^2$ are radial and
tangential components of velocity dispersion respectively ($0\le A < 2$). In general
case it depends on radial distance and time. The aim of this work is to determine
the value of anisotropy parameter for 10 globular clusters, to try find out the its
dependence on limiting magnitude and on radial distance.

\section{The Method}

We have used the method that was proposed by Agekian and Petrovskaya (1962).
Let us assume: 1) the star ensemble has an equal-mass spectrum; 2) the cluster
is spherically symmetric and is in stationary state in the regular force-field;
3) the cluster consists of single stars; 4) $\sigma_r^2=const$ and $\sigma_t^2=const$
along radius (so $A=const$). Then the cluster is described by means of the equation
$$
\displaystyle \frac {d}{d\xi} \left(\displaystyle \frac {\xi^2}{D}\displaystyle
\frac {dD}{d\xi} \right)=-D \xi^2-A,\eqno (1)
$$
where $\xi$ and $D$ are undimensional radial distance and undimensional spatial density.
A theoretical surface mass density profile is determined as
$$
F_{theor}(r)=-2 \beta \int\limits_{r/\alpha}^{\infty}
\sqrt{\xi^2-\left( r/\alpha \right)^2} D'\left(\xi\right)d\xi-F_0,\eqno (2)
$$
where $\alpha$ and $\beta$ are scale factors, $F_0$ is constant taking into account
the infinity radius of the model and $D'\left(\xi\right)$ is the solution
of (1). From assumption 1) it follows that $\rho\propto n$, where n is spatial number
star density. Then $F_{theor}(r)$ is theoretical surface number density of
stars. Here constants $\alpha$, $\beta$ and $F_0$ are determined by minimization
of the function $\Phi(A)=\sum \limits_{i=1}^N\left[F_{theor}(r_i)-F_{obs}(r_i)\right]^2$
where N is number of star counts zones and $F_{obs}(r)$ is observed surface number
density of stars. That value of $A$ that gives minimum $\Phi(A)$ is adopted as most
probable value of A.

\section{Results and Discussion}

We determined the value of anisotropy parameter $A$ for following 10 globular clusters:
M3, M12, M13, M15, M55, M56, Palomar 3, Palomar 14, NGC5139 ($\omega$ Cen) and
NGC6535. The results are given in Table 1. For most clusters we have $A=0$,
i.e. they have almost isotropic velocity distribution. It is interesting that for all
clusters theoretical number density profiles are in good agreement with observational
ones. But the theoretical density profile in the halo almost in all cases is lower than
observational one. Perhaps the cause is in the assumption about equal-mass. The fact is,
as a result of mass segregation the low-mass stars prevail in the halo and consequently
we overestimated mass density in this region. In reality the mass density less than we suppose.

For M56 at limiting magnitude $B=21\fm 5$ we found $A=0.8$. Hence in this cluster
the velocity distribution is strongly anisotropic yet. Besides in order to find out an influence
of limiting magnitude to value of $A$, we calculated it at $B=19\fm 0$ and found
$A=0.3$, i.e. if in this case the limiting magnitude decreases, then the anisotropy parameter
decreases also. Perhaps it is because, in first case ($B=21\fm 5$) it is expected the
influence of low-mass stars to value of $A$ that generally are in halo and move along
elongated orbits.

In order to find out the possible dependence of the anisotropy parameter on
the radial distance we performed for NGC5139 ($\omega$ Cen) additional calculations in
three separate regions. The results show that in the central region $A=0$, in middle
region $A=0.1$ and in the halo $A=0.3$. In this case the agreement of the theoretical
and observational density profiles essentially was improved than one for whole cluster.

\begin{table}
\caption{Value of anisotropy parameter for 10 globular clusters.}
\begin{tabular}{ccccc}
\tableline
Cluster & A & Limit. magnit. & Source of         & Note\\
        &   & or Exp. time  & observational data &      \\
\tableline
M3      & 0   & 10 min     & King et al. 1968 &      \\
M12     & 0   &  9 min     & King et al. 1968 &      \\
M13     & 0   & 10 min     & King et al. 1968 &      \\
M15     & 0   & 30 min     & King et al. 1968 &      \\
M55     & 0   &  2 min     & King et al. 1968 &      \\
Pal 3   & 0   & 30 min     & King et al. 1968 &      \\
Pal 14  & 0   & V=22\fm 4  & Harris \&        &      \\
        &     &            & van den Berg 1984 &     \\
NGC6535 & 0   & B=20\fm 0  & Peykov \&         &      \\
        &     &            & Roussev 1988      &      \\
M56     & 0.8 & B=21\fm 5 & Peykov \&          &      \\
        &     &           & Roussev 1986       &      \\
 -//-    & 0.3 & B=19\fm 0 & Peykov \&         &      \\
        &     &           & Roussev 1986       &      \\
NGC5139 & 0   & 30 min & King et al. 1968 & $6.27\le r\arcmin \le 23.82$ \\
 -//-   & 0   &  -//-  & King et al. 1968 & $6.27\le r\arcmin \le 10.22$ \\
 -//-   & 0.1 & -//-   & King et al. 1968 & $10.22 < r\arcmin \le 14.75$ \\
 -//-   & 0.3 & -//-   & King et al. 1968 & $14.75 < r\arcmin \le 23.82$ \\
\tableline
\tableline
\end{tabular}
\end{table}

We think this way allowes to solve partly the problem of dependence of the
anisotropy parameter on radial distance for any globular cluster. If we will make use
of the more precise observational data, e.g. surface brightness profile recevied from
Hubble Space Telescope, then we can find the more precisely dependence of anisotropy
parameter on radial distance in the concrete observed globular cluster. Further we plan
to study nostationary effects.


\begin{references}
\reference

Agekian, T. A., \& Petrovskaya, I. V. 1962, Uchen.zap.LGU, 307, 187

Bettwieser, E., \& Spurzem, R. 1986, \aap, 161, 102

Harris, W. E., \& van den Berg, S. 1984, \aj, 89, 1816

King, I. R., Hedemann, E. Jr., Hodge, S.M., \& White, R. E. 1968, \aj, 73, 456

Louis, P. D. 1990, \mnras, 244, 478

Meylan, G., \& Heggie, D. C. 1997, \aap Rew, 8, 31

Peykov, Z. I., \& Roussev, R. M. 1986, \azh, 63, 483

Peykov, Z. I., \& Roussev, R. M. 1988, \azh, 65, 317

Solanes, J. M., \& Salvador-Sole, E. 1990, \aap, 234, 93

Spurzem, R. 1996, in IAU Symp. 174, Dynamical Evolution of Star
Clusters - Confrontation of Theory and Observations, eds. P.Hut \& J.Makino
(Dordrecht: KAP), 111

Takahashi, K. 1996, in IAU Symp. 174, Dynamical Evolution of Star
Clusters - Confrontation of Theory and Observations, eds. P.Hut \& J.Makino
(Dordrecht: KAP), 91

\end{references}
\end{document}